\documentclass{ws-procs9x6}

\begin{document}
\title{Parameterized post-Newtonian limit\\of Horndeski's gravity theory}

\author{M. Hohmann$^*$}

\address{Laboratory of Theoretical Physics, Institute of Physics, University of Tartu,\\
Tartu, Tartumaa, Estonia\\
$^*$E-mail: manuel.hohmann@ut.ee\\
http://kodu.ut.ee/\textasciitilde{}manuel/}

\begin{abstract}
We present a recent result on the parameterized post-Newtonian (PPN) limit of Horndeski's gravity theory and its consistency with solar system observations.
\end{abstract}


\bodymatter

\section{Method}\label{sec:method}
The starting point of our calculation are the most general second order field equations of scalar-tensor gravity,\cite{Kobayashi:2011nu} whose dynamical variables are the metric \(g_{\mu\nu}\) and the scalar field \(\phi\) and whose action contains four free functions \(K, G_3, G_4, G_5\) of \(\phi\) and its kinetic term \(X = -\nabla_{\mu}\phi\nabla^{\mu}\phi/2\). We expand the fields \(g_{\mu\nu} = \eta_{\mu\nu} + h_{\mu\nu}\) and \(\phi = \Phi + \psi\) around their cosmological background values \(\eta_{\mu\nu}, \Phi\). Further, we expand the free functions in a Taylor series
\begin{equation}\label{eqn:taylorseries}
K(\phi,X) = \sum_{m,n = 0}^{\infty}K_{(m,n)}\psi^mX^n\,,
\end{equation}
and analogously for \(G_3,G_4,G_5\). We then introduce post-Newtonian orders of magnitude for the field perturbations \(h_{\mu\nu}, \psi\) and expand the field equations in these orders. We finally solve the field equations order by order.

\section{Results}\label{sec:results}
We now display the results obtained from applying the method shown in the previous section to Horndeski's gravity theory.\cite{Hohmann:2015kra} It turns out that the only PPN parameters which potentially deviate from their observed values are \(\gamma\) and \(\beta\). In general they depend on the distance \(r\) between the gravitating mass and the test mass and take the form
\begin{eqnarray}\label{eqn:gammabeta}
\gamma(r) &=& \frac{2\omega + 3 - e^{-m_{\psi}r}}{2\omega + 3 + e^{-m_{\psi}r}}\,,\\
\beta(r) &=& 1 + \frac{1}{(2\omega + 3 + e^{-m_{\psi}r})^2}\Bigg\{\frac{\omega + \tau - 4\omega\sigma}{2\omega + 3}e^{-2m_{\psi}r}\nonumber\\
&&+ a(r)\left[e^{-m_{\psi}r}\ln(m_{\psi}r) - \left(m_{\psi}r + e^{m_{\psi}r}\right)\mathrm{Ei}(-2m_{\psi}r) - \frac{1}{2}e^{-2m_{\psi}r}\right]\nonumber\\
&&+ b(r)\left[e^{m_{\psi}r}\mathrm{Ei}(-3m_{\psi}r) - e^{-m_{\psi}r}\mathrm{Ei}(-m_{\psi}r)\right]\Bigg\}\,,
\end{eqnarray}
where we used the abbreviations
\begin{equation}
a(r) = (2\omega + 3)m_{\psi}r\,, \quad b(r) = \frac{6\mu r + 3(3\omega + \tau + 6\sigma + 3)m_{\psi}^2r}{2(2\omega + 3)m_{\psi}}\,,
\end{equation}
and introduced the constants
\begin{gather}
m_{\psi} = \sqrt{\frac{-2K_{(2,0)}}{K_{(0,1)} - 2G_{3(1,0)} + 3\frac{G_{4(1,0)}^2}{G_{4(0,0)}}}}\,, \quad \omega = \frac{G_{4(0,0)}}{2G_{4(1,0)}^2}\left(K_{(0,1)} - 2G_{3(1,0)}\right)\,,\nonumber\\
\sigma = \frac{G_{4(0,0)}G_{4(2,0)}}{G_{4(1,0)}^2}\,, \quad \tau = \frac{G_{4(0,0)}^2}{2G_{4(1,0)}^3}(K_{(1,1)} - 4G_{3(2,0)})\,, \quad \mu = \frac{G_{4(0,0)}^2K_{(3,0)}}{G_{4(1,0)}^3}\,.
\end{gather}
In the limit \(m_{\psi} \to 0\) of a massless scalar field we obtain the values
\begin{equation}
\gamma = \frac{\omega + 1}{\omega + 2}\,, \quad \beta = 1 + \frac{\omega + \tau - 4\omega\sigma}{4(\omega + 2)^2(2\omega + 3)}\,,
\end{equation}
which are independent of the interaction distance \(r\).

\section*{Acknowledgments}
The author gratefully acknowledges the full financial support of the Estonian Research Council through the Postdoctoral Research Grant ERMOS115 and the Startup Research Grant PUT790.

\end{document}